%% file: main_rv3.tex
\documentclass[%
reprint,
superscriptaddress,
amsmath,amssymb,
aps,
pre,
colorlinks
]{revtex4-2}

\usepackage{graphicx}
\usepackage{dcolumn}
\usepackage{bm}
\usepackage[usenames,dvipsnames]{xcolor}
\usepackage[urlcolor=NavyBlue,citecolor=NavyBlue,linkcolor=NavyBlue]{hyperref}

\usepackage{physics}   
\usepackage{bm}		   
\usepackage{mathtools}
\usepackage{bbold}     
\usepackage[]{changes}   
\usepackage{lineno}     
\usepackage{xr}

\newcommand{\vbf}{\boldsymbol{v}}
\newcommand{\Vbf}{\boldsymbol{V}}
\newcommand{\fbf}{\boldsymbol{f}}
\newcommand{\xibf}{\boldsymbol{\xi}}

\newcommand{\zetabf}{\boldsymbol{\zeta}}
\newcommand{\etabf}{\boldsymbol{\eta}}
\newcommand{\thetabf}{\boldsymbol{\theta}}
\newcommand{\xbf}{\boldsymbol{x}}

\newcommand{\defeq}{\mathrel{\mathop:}=}
\newcommand{\eqdef}{=\mathrel{\mathop:}}
\let\originalleft\left
\let\originalright\right
\newcommand{\mleft}{\mathopen{}\mathclose\bgroup\originalleft}
\newcommand{\mright}{\aftergroup\egroup\originalright}
\newcommand{\ddint}[1]{\!\dd{#1}}
\newcommand{\ie}{i.e.\@}

\newcommand{\subheading}[1]{\vspace{3mm}
	\noindent
	\textbf{#1}\newline}
\renewcommand{\vec}[1]{\boldsymbol{ #1 }}
\renewcommand{\expval}[1]{\mathbb{E}\mleft[ #1 \mright]}

\graphicspath{{./figs/}}

\begin{document}
	\preprint{APS/123-QED}
	
	\title{A Likelihood Approach for Inference of Population Heterogeneity in Particle Ensembles with Second-Order Langevin Dynamics}
	
	\author{Jan Albrecht}
	\email{jan.albrecht@uni-potsdam.de}
	\affiliation{Institute of Physics and Astronomy, University of Potsdam, 14476 Potsdam, Germany}
 	
	\author{Manfred Opper}
	\email{manfred.opper@tu-berlin.de}
	\affiliation{Faculty of Electrical Engineering and Computer Science, Technische Universit\"at Berlin, 10587 Berlin, Germany}

	\author{Robert Gro{\ss}mann}%
	\email{rgrossmann@uni-potsdam.de}
	\affiliation{Institute of Physics and Astronomy, University of Potsdam, 14476 Potsdam, Germany}
 
	\date{\today}

	\begin{abstract}
		\input{ABS}
	\end{abstract}
	
	\maketitle	
	
	\section*{Introduction}
	\input{INTRO}

	\section*{Results}
	\input{MODEL}

	\input{APPROACH_NEW}

	\subheading{Failure of naive likelihood approximations for second order SDEs}
	\input{LINEAR}

	\subheading{Likelihood approximation for second-order SDEs}
	\input{TR_GAU}

	\subheading{Likelihood maximization via the EM algorithm}
	\input{EM}

	\subheading{Numerical investigation of heterogeneity estimation}
	\input{RES_CONST}

	\section*{Discussion}
	\input{CONC}

	\section*{Methods}
	\input{HESS_MAT}

\input{APP}

	\input{ACKN}
	\nolinenumbers
	
	\bibliography{het_like_rg}	
	
\end{document}

%% file: ABS.tex
The inherent complexity of biological agents often leads to motility behavior that appears to have random components. Robust stochastic inference methods are therefore required to understand and predict the motion patterns from time-discrete trajectory data provided by experiments. In many cases, second-order Langevin models are needed to adequately capture the motility. Additionally, population heterogeneity needs to be taken into account when analyzing data from several individual organisms. In this work, we describe a maximum likelihood approach to infer dynamical, stochastic models and, simultaneously, estimate the heterogeneity in a population of motile active particles from discretely sampled, stochastic trajectories. To this end, we propose a method to approximate the likelihood for non-linear second-order Langevin models. We show that this maximum likelihood ansatz outperforms alternative approaches, especially for short trajectories. Additionally, we demonstrate how a measure of uncertainty for the heterogeneity estimate can be derived. We thereby pave the way for the systematic, data-driven inference of dynamical models for actively driven entities based on trajectory data, deciphering temporal fluctuations and inter-particle variability.

%% file: INTRO.tex
One of the drivers of active matter research has been the goal to understand and describe the motility behavior of biological agents which span orders of magnitude from microorganisms and cancer cells to flocks of sheep and birds~\cite{vicsek2012collective, Romanczuk2012, Marchetti2013, mehes_collective_2014, Bechinger2016}. The inherent complexity of even single-cellular organisms and the plethora of unobserved internal degrees of freedom leads to motility patterns which appear to have random components. For this reason, stochastic models and, more specifically, stochastic differential equations~(SDEs) or Langevin equations, have emerged as an important model class to describe the motility of macroscopic~\cite{Cavagna2015, 
	Gautrais2012, 
	Ariel2015, 
	Corbetta2017} 
as well as microscopic biological agents~\cite{Selmeczi2008, AmselemBeta2012, PedersenFlyvbjerg2016, KlimekNetz2024, Bruckner2024}. 

Motility data from active particles takes the form of discretely sampled trajectories. 
In experimental setups it is often not possible to follow a single individual for an extended amount of time: the particles can move out of the frame \cite{Moldenhawer2022} or plane of focus \cite{Alirezaeizanjani2020}, or become obscured by other particles moving in front of them \cite{Cavagna2013}.
Instead of a few long trajectories, only short trajectories from many different individuals are available in these situations.
This poses a challenge especially if the observed population is not homogeneous, \ie, there is inter-individual variability. While individuality of macroscopic animals might be intuitive,  even populations of genetically identical microscopic organisms display heterogeneity~\cite{Meacham2013, Peled2021, Persson2022, Ariel2022}. This population heterogeneity can manifest itself in the motility behavior of the organisms~\cite{Spudich1976, Waite2016}. Even if the motility of all particles is captured by the same type of motility model, the parameters of this model will differ from individual to individual in heterogeneous populations. Population heterogeneity can lead to unusual behavior of statistical quantities, e.g.~non-Gaussian displacement distributions~\cite{Lemaitre2023, Grossmann2024}. Ignorance about the heterogeneity can therefore result in wrong conclusions about the model~\cite{Metzler2020}.

So far, heterogeneity inference for populations of motile microorganisms has mostly been limited to splitting the whole population into several subpopulations according to properties of the individual trajectories~\cite{AmselemBeta2012, Bruckner2020}. 
Other approaches to infer population heterogeneity in biological systems include parameter fitting for each individual particle~\cite{Grossmann2024} and using forward simulations of the system to optimize a superposition of ansatz functions~\cite{Hasenauer2010MLE, Hasenauer2011Analysis}. Recently, it has also been suggested to use forward simulations to determine the parameter spread within a population from trajectory data~\cite{Klimek2025}. 
A general approach that enables one to infer arbitrarily parametrized continuous distributions and that utilizes the, possibly limited, trajectory information in an optimal way through likelihood based methods is still missing. Likelihood-based methods also allow uncertainty quantification of the estimates and still work in situations with limited data.

Similar challenges have emerged in biomedical research in general and in the subfield of pharmacokinetics and pharmacodynamics in particular~\cite{Donnet2013}. The metabolism of medical drugs can be described by SDEs, but the parameters of these SDEs may vary significantly between individual patients. In this community such SDEs with heterogeneity of the parameters are known as stochastic differential mixed effects models~(SDMEMs). A number of methods and algorithms have been proposed to solve this inference problem.  They range from analytical approximations~\cite{Picchini2010} over stochastic approximations~\cite{DelattreLavielle2013, Baltazarlarios2024} to Bayesian inference using sophisticated particle methods~\cite{Whitaker2017, Wiqvist2021}; we point to Ref.~\cite{PicchiniList} for an extensive list of references. There are also a number of contributions about theoretical properties of estimators in such systems~\cite{Delattre2021}. However, there are some peculiarities about motility data that call for specialized methods as detailed in the following.

First, the positional dynamics of active particles will often require a second-order description. In macroscopic inertial systems such as flocks of birds, second-order dynamics arises naturally through the Newtonian equations of motion~\cite{Cavagna2015}. But also for microorganisms whose motion is usually governed by low Reynolds numbers and is therefore overdamped, a first-order description might not be sufficient.  
Since the propulsion mechanism of motile cells, for example, generates fluctuation forces and thus fluctuating velocities, it has become common to model these velocities by first-order Langevin equations \cite{PedersenFlyvbjerg2016, Bodeker2010, Bruckner2019}, which leads to a second-order equation in position.
Typically, the full dynamics is only partially observable, as the dynamics of the propulsion mechanism is not experientally accessible---oftentimes, only the positions can be recorded. Due to the underlying second-order dynamics, the observed positional process is non-Markovian and does therefore not allow for parameter inference through transition density estimation. 
Recently, new inference methods for homogeneous second-order motility models have been introduced~\cite{BrucknerBroedersz2020, Ferretti2020}. However, these studies focus on the calculation of single point estimates for the motility parameters of each individual. In order to estimate the population heterogeneity from the available data in an optimal way, the full likelihood of the motility parameters needs to be taken into account. This is especially important if the measured trajectories have a short duration limiting the available information per trajectory. 
In the context of discrete time processes instead of continuous SDEs, motility inference based on likelihoods was discussed in Ref.~\cite{metzner2015superstatistical}---the parameters inferred by this method will, however, depend on the sampling interval. 
A methodologically different approximation of the likelihood in second-order settings can be obtained using extended Kalman Filters~\cite{DelattreLavielle2013, Sarkka2013}. These provide a general approximation for partially observed systems by sequentially integrating out the unobserved parts of the system. This sequential algorithm, however, does not provide a closed-form expression for the likelihood and cannot provide intuitions about its parameter dependence. 

Second, while individual trajectories may be very short, many experimental settings allow, however, for high sampling rates. This means that the relevant limit in this case is short time steps~$\tau$ while the total duration of the trajectories~$T$ remains constant. An inference method for motility data needs to be able to deal with potentially many data points per trajectory while converging to the true likelihood in the small~$\tau$ limit. 

The main result of this work is to address the challenges summarized above by presenting and discussing a maximum likelihood based framework to infer distributions of motility parameters in heterogeneous ensembles of active particles. 
Maximum likelihood estimators~(MLEs) are known to have favorable theoretical properties like consistency, equivariance and efficiency. To this end, we introduce an approximate expression for the likelihood of the parameters of the motility model with respect to single trajectories in non-linear second-order settings. This method yields an analytical approximation to the likelihood with a concise functional form. The derivation has parallels with the approach taken in Ref.~\cite{Ferretti2020} which considers a smaller model class.
To obtain an MLE for the heterogeneity, we plug the likelihood approximation into an expectation maximization~(EM) scheme. We evaluate the performance of the approach using numerical data and show how the method provides robust confidence intervals.

%% file: MODEL.tex
\subheading{A model for heterogeneous particle ensembles}
We aim to describe a population of active particles with inter-particle variability, \ie, a heterogeneous particle ensemble. Our main modeling assumption is that all particles have the basic same propulsion machinery that is captured by a certain type of motility model. However, we also assume that inter-individual variability of the particles leads to randomness in the tuning of this machinery and, thus, to randomness in the parameters of the motility model. To capture this we employ a hierarchical model. 

First, we model the dynamics of individual particles, \ie, their motility. In this work, we focus on the case in which the motility is described by a Langevin equation in velocity with additive noise of the form
\begin{gather}
	\ddot{\vec{x}}^n(t) = 
	\dot{\vbf}^n(t) = \fbf(\vbf^n(t); \etabf^n_f) + \sqrt{2 D^n}\,\xibf^n(t)\,, 
	\label{eq:motility_model}
\end{gather}
where the uppercase index $n$ labels the specific particle. Here, $\fbf$ is a smooth~$\mathbb{R}^d \times \mathbb{R}^{q-1}\rightarrow\mathbb{R}^d$ function and~$\xibf^n(t)$ are~$d$-dimensional white noise processes with~$\expval{\vec{\xi}^n(t)(\vec{\xi}^{m})^T(t')} = \mathbb{1}\,\delta_{nm}\delta(t - t')$.
The motility model is parametrized by \textit{motility parameters} $\etabf^n=\{\etabf_f^n, D^n\}\in \mathbb{R}^q$. For applicability to cell motion see Supplementary Note 7. Note that we use the term \textit{Langevin equation} for general SDEs in accordance with popular textbooks like Ref.~\cite{Risken1996}.

As a second part we assume that the distribution of these motility parameters in the population follows a certain probability distribution
\begin{gather}
	\etabf^n\sim p_{\etabf}(\cdot |\thetabf^*)\quad \mathrm{i.i.d.}\,.
\end{gather}
This heterogeneity distribution is parameterized by \textit{heterogeneity parameters} $\thetabf \in \mathbb{R}^r$. We denote the true heterogeneity parameters~$\thetabf^*$.

In an experiment, the positions of~$N$ particles are observed at fixed time intervals~$\tau$ so that we measure a trajectory~$\boldsymbol{T}^n = \left\{\xbf_j^n\right\}_{j = 0, 1, \ldots, M^n}\eqdef \xbf^n_{0:M^n}$ for each particle~$n$, where the~$j^\mathrm{th}$ position is recorded at time~$t^n_j = t^n_0 + j\tau$. The number of measurement points in trajectory~$n$ is then~$M^n+1$, so that~$M^n$ is the number of measured displacements. The duration of a trajectory is then~$T^n \defeq t_{M^n}^n - t_0^n = M^n\,\tau$. 

We use $\mathcal{D}\defeq \left\{\vec{T}^n\right\}_{n=1,\ldots,N}$ to denote the full set of trajectories.

%% file: APPROACH_NEW.tex
\subheading{Likelihood-based approach for heterogeneity estimation}
\begin{figure*}[tb]
	\begin{centering}
		\includegraphics[width=\textwidth]{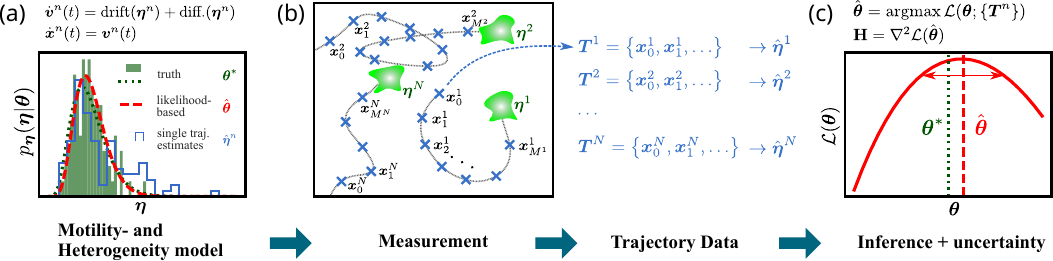}	
	\end{centering}
	\vspace*{-0.7cm}
	\caption{\textbf{Schematic sketch of heterogeneity inference.} (a) A stochastic motility model in velocity is assumed as well as a distribution~$p_{\boldsymbol{\eta}}$ of the motility parameters~$\etabf$. This heterogeneity distribution has some true parameter~$\thetabf^*$. (b) Each observed active particle moves with a given set of motility parameters~$\vec{\eta}^n$. Discretely sampled trajectories~$\boldsymbol{T}^n$ of the particles are measured. For each of these trajectories an estimate of the motility parameter $\hat{\vec{\eta}}^n$ can be calculated. (c) The likelihood of the heterogeneity parameters with respect to all measured trajectories is used to obtain an estimate~$\hat{\thetabf}$. The curvature of the likelihood is a measure of uncertainty for the estimate. As can be seen in (a), the likelihood estimation recovers the true heterogeneity while the distribution of single trajectory estimates does not capture the true input distribution.
	Plots (a) and (c) show results for inference on an integrated Ornstein-Uhlenbeck process with heterogeneous mean reversion rate.
}
	\label{fig:model_drawing}
\end{figure*}

To characterize the observed system, we want to infer the heterogeneity within the population of particles from the observed discrete trajectories; the whole process from model over measurement to inference is shown schematically in Fig.~\ref{fig:model_drawing}. 

Our main goal is to obtain an accurate estimate~$\hat{\thetabf}$ of the true heterogeneity parameters~$\thetabf^*$ from the observed dataset of trajectories $\mathcal{D}$. Here and in the remainder of the text, we use hats to denote estimates of quantities.
A straight forward approach would be to first analyze every trajectory $\vec{T}^n$ individually and calculate an estimate $\hat{\vec{\eta}}^n$ for its motility parameter. The distribution of these estimates in parameter space can then be used to estimate the heterogeneity of the system \cite{Delattre2015, GenonCatalot2016, Grossmann2024}. We call this a two-step approach. However, this approach ignores the uncertainty attached to each of the single trajectory estimates $\hat{\vec{\eta}}^n$  due the stochasticity of the trajectories. Especially in situations where this uncertainty is large the two-step approach fails to recover the true heterogeneity of the system. Without additional tests, it is also unclear if the spread of the estimated parameters $\hat{\vec{\eta}}^n$ reflects underlying heterogeneity or simply arises from finite-size fluctuations of the stochastic trajectories. A recent publication discusses this issue for the special case of anomalously diffusing particles \cite{Lanoiselee2025SuperResolved}.

We, therefore, propose here a full likelihood-based approach to estimate the heterogeneity. This approach allows to go directly from the trajectory data to the heterogeneity estimate and implicitly takes the uncertainty about motility parameters into account. 

The log-likelihood of the heterogeneity parameters $\mathcal{L}(\thetabf) \defeq \log p(\mathcal{D}|\thetabf)$ is just the log-probability of the observed trajectory data conditioned on the heterogeneity parameters. The value $\hat{\thetabf}$ that maximizes this function is the maximum likelihood estimator (MLE) of $\vec{\theta}$. The log-likelihood can be expressed in the following way:
\begin{align}
	\mathcal{L}(\thetabf) 
	& = \sum_n \log p(\boldsymbol{T}^n|\thetabf)=\sum_n\log \int\ddint{\etabf^n} p(\boldsymbol{T}^n, \etabf^n|\thetabf)\nonumber\\
	&=\sum_n\log \int\ddint{\etabf^n} p(\boldsymbol{T}^n|\etabf^n)\, p_{\etabf}(\etabf^n|\thetabf)\,.
	\label{eq:MLE_likelihood}
\end{align}
Under the integral in Eq.~\eqref{eq:MLE_likelihood} another likelihood expression appears: the likelihood of the motility parameters $p(\vec{T}^n|\vec{\eta}^n)$. This term encodes the uncertainty about the true motility parameter $\etabf^n$. The full likelihood approach also provides an uncertainty estimate of $\hat{\thetabf}$ through the curvature of $\mathcal{L}(\thetabf)$ at its maximum \cite{Schervish1995}, 
\begin{gather}
	H_{\alpha, \beta} = \left.\pdv{\theta^{\alpha}}\pdv{\theta^{\beta}}\mathcal{L}(\thetabf) \right|_{\thetabf = \hat{\thetabf}}\,,
	\label{eq:MLE_hessianmatrix}
\end{gather}
where upper Greek indices denote components of vectors. With this, it can be assessed whether the estimated heterogeneity is indeed significant.

There are two challenges when trying to apply the likelihood approach. First, an approximation to the usually untractable log-likelihood for the motility parameters~$L_n(\eta) = \log p(\boldsymbol{T}^n|\etabf)$ needs to be found. We call these likelihoods \textit{single trajectory log-likelihoods}. While~$p_{\etabf}(\cdot|\thetabf)$ is given from the model, the single trajectory likelihood~$p(\boldsymbol{T}|\etabf)$ needs to be derived from the motility model~[cf.~Eq.~\eqref{eq:motility_model}]. The second-order nature of the model requires to go beyond transition density estimates which are often employed for first-order SDEs. Second, we require a structured way to maximize the log-likelihood $\mathcal{L}(\thetabf)$ that can deal with the intractable integrals in Eq.~\eqref{eq:MLE_likelihood}. 

In the following, we show how to solve these challenges. We first discuss and derive approximations of the single trajectory likelihood. After that we show how an EM algorithm can be used to maximize the likelihood and test the framework on numerically generated data.

%% file: LINEAR.tex
For SDEs of first order, there are plenty of methods which provide parameter estimators as well as approximated likelihood expressions~\cite{Iacus2008}. In this paper, we consider cases where only the integral of a first-order process in velocity is observed which leads to a second-order model in position. The instantaneous velocity process can be approximated from positional data by using finite differencing
\begin{align}
	\Vbf_j = \frac{1}{\tau} \int_{t_j}^{t_{j+1}} \ddint{t} \vec{v}(t) = \frac{\vec{x}_{j+1} - \vec{x}_j}{\tau}\,.
	\label{eq:linproc_secantvelocities}
\end{align}
Following Ref.~\cite{PedersenFlyvbjerg2016}, we will refer to~$\Vbf_j$ as \textit{secant velocities} in contrast to the instantaneous velocities~$\vec{v}_j \defeq\vec{v}(t_j)$. 
In the fast sampling limit of~$\tau\rightarrow 0$, the approximation~$\vec{v}_j \approx \Vbf_j$ becomes exact.
However, using estimation techniques for the first-order process and plugging in the discrete approximation will lead to biases that persist even in the limit of~$\tau\rightarrow0$. Supplementing the discussions in Refs.~\cite{Bruckner2020, Ferretti2020, CialencoPasemann2024, Albrecht2026}, we want to give here a clear and concise explanation how and why this bias arises. 
Intermediate steps and details on the calculations in this sections can be found in Supplementary Note 1.

As an example, we consider a model in which the velocity dynamics is described by the linear Ornstein-Uhlenbeck~(OU) process:
\begin{gather}
	\dot{\vbf}(t) = -\gamma^* \vbf(t) + \sqrt{2D^*}\, \xibf (t) \,.
	\label{eq:linproc_OUP}
\end{gather}
Using an Euler-style approximation to the log-likelihood with respect to direct observations of the velocity~$\left\{\vec{v}_j\right\}$, we can obtain 
the following MLEs for the two motility parameters
\begin{subequations}
	\label{eq:linproc_vMLE}
	\begin{align}
		\hat{\gamma}_{v, \mathrm{Eul}} &= -\frac{\frac{1}{M_{\vec{v}}}\sum_j \vbf_j\Delta \vbf_j }{\tau\frac{1}{M_{\vec{v}}}\sum_k{\vbf_k^2}}\,,\\
		\hat{D}_{v, \mathrm{Eul}} &= \frac{1}{2\,d\,\tau} \, \frac{1}{M_{\vec{v}}}\sum_j\left(\Delta \vbf_j + \hat{\gamma}_{v, \mathrm{Eul}} \vbf_j \tau\right)^2\,. 
	\end{align}
\end{subequations} 
Here, $M_{\vec{v}}$ is the number of terms in each of the sums and $\Delta \vbf_j \defeq \vbf_{j+1} - \vbf_j$.
The likelihood as well as these estimators will become accurate in the limit of~$\tau\rightarrow 0$. In this limit the correlation between the estimates will also vanish since then $\Delta \vbf_j\gg\hat{\gamma}_{v, \mathrm{Eul}} \vbf_j \tau$. If the instantaneous velocities cannot be observed but only positions, a naive approach would be to plug the finite difference approximation from Eq.~\eqref{eq:linproc_secantvelocities} into the estimators~\eqref{eq:linproc_vMLE}. The OU process is ergodic and we can therefore replace the sums with expectation values for large $M_{\vec{v}}$. This leads to the following estimators:
\begin{subequations}
	\begin{align}
		\hat{\gamma}_{\mathrm{Eul}} &\approx -\frac{\mathbb{E}\mleft[\vec{V}_j\Delta\vec{V}_{j}\mright] }{\tau\mathbb{E}\mleft[{\Vbf_j}^2\mright]} , \\
		\hat{D}_{\mathrm{Eul}} &\approx \frac{1}{2\,d\,\tau} \, \mathbb{E}\mleft[\left(\Delta \Vbf_j + \hat{\gamma}_{\mathrm{Eul}} \Vbf_j \tau\right)^2\mright]\,. 
	\end{align}
\end{subequations}
Plugging in the true expressions for the covariances of~$\vec{V}_j$~(see Supplementary Note 1) and taking the limit~$\tau\rightarrow 0$, we obtain
\begin{gather}	
\label{eqn:biasesOUP}\hat{\gamma}_{\mathrm{Eul}}\rightarrow \frac{2}{3}\gamma^*, \qquad \hat{D}_{\mathrm{Eul}}\rightarrow \frac{2}{3} D^*\,.
\end{gather}
This means that the ML estimate is biased even in the limit of high frequency sampling and long trajectories.
The bias factor of~$2/3$ is typical for inference from integrals of processes driven by Brownian motion~\cite{CialencoPasemann2024, Gloter2006, Ferretti2022, Albrecht2026}.

We now take a step back from the above concrete linear example and discuss on a more general and abstract level why naive approaches fail. 
Above, we demonstrated that maximum likelihood estimators include expectation terms of the form~$\expval{\Delta \vec{V}_j\Delta \vec{V}_j}$ and $\expval{\vec{V}_j\Delta \vec{V}_j}$.
These terms depends on correlation structures on a time scale of~$\tau$. 
On long timescales, the statistics of the instantaneous and secant velocity become equivalent, since~$\vec{V}$ converges to~$\vec{v}$.
However, on small timescales comparable to~$\tau$, the statistics and correlation structure of~$\vec{v}$ and~$\vec{V}$ differ which leads to estimation biases as in Eq.~\eqref{eqn:biasesOUP}.

We consider the one-dimensional case for simplicity. In the stationary case, the following relation holds: 
\begin{align}
	\expval{V_j\Delta V_j} =& \mathbb{E}[V(0) V(\tau)] -\mathbb{E}[V(0)V(0)]\,.
	\label{eq:fail_VDV}
\end{align}
With the definition of the secant velocity in Eq.~\eqref{eq:linproc_secantvelocities}, correlations of the secant velocity can be expressed as 
\begin{gather}
	\mathbb{E}[V(0)V(j\tau)] = \frac{1}{\tau^2}\int_0^\tau\ddint{t'}\int_{j\tau}^{(j+1)\tau}\ddint{t''} \mathbb{E}[v(t')\, v(t'')]\,.
	\label{eq:fail_EVV}
\end{gather}
If the velocity dynamics is a stationary diffusion process, the autocorrelation function is well approximated by 
$\mathbb{E}[v(0)\, v(t)]=c_0 + c_1 \abs{t} +\ldots\,$ in a small region around zero with $c_0 = \expval{v^2}$  and $c_1 = \expval{v\,f(v)}$. Importantly, $c_1\neq 0$ which implies that the derivative of the autocorrelation function is discontinuous at the origin. Using the above expansion, we find 
\begin{subequations}
	\begin{align}
		\expval{V(0)V(0)} &= c_0 + \frac{1}{3} c_1 \tau + \mathcal{O}(\tau^2) ,\\
		\expval{V(0)V(\tau)} &= c_0 + c_1 \tau + \mathcal{O}(\tau^2)\,.
	\end{align}	
	\label{eq:fail_corrV}
\end{subequations}
Plugging these results into Eq.~\eqref{eq:fail_VDV}, we eventually obtain
\begin{gather}
	\mathbb{E}[V_j \Delta V_j] = \frac{2}{3}c_1\tau = \mathbb{E}[v_j \Delta v_j]\left(\frac{2}{3} +\mathcal{O}(\tau)\right). 
	\label{eq:fail_biasfactor}
\end{gather}
Similarly, we find that 
\begin{gather}
	\expval{\Delta V_j\Delta V_j} = \mathbb{E}[\Delta v_j \Delta v_j]\left(\frac{2}{3} +\mathcal{O}(\tau)\right)\,.
\end{gather}
The bias factor of~$2/3$ here is the same factor that appeared in Eq.~\eqref{eqn:biasesOUP}. 

Note that this result is connected to the non-differentiability of the velocity autocorrelation function at zero. It is well known for Gaussian processes that the behavior of the autocorrelation at zero is linked to the roughness of the process~\cite{Rasmussen2005}. 
Therefore the bias arises due to the roughness of the underlying velocity process, which is driven by Gaussian white noise. 

In the following section we will present a likelihood approximation for second-order SDEs that avoids these biases.

%% file: TR_GAU.tex
\label{sec:trGau}
In this paper, we consider motility models of the form 
\begin{gather}
	\dot{\vbf}(t) = \fbf(\vbf(t); \etabf_f) + \sqrt{2 D}\,\xibf(t)\,.
	\label{eq:trgau_SDEv}
\end{gather}
with non-linear drift $\vec{f}(\vec{v})$. For these, there is no general analytic expression for the single trajectory likelihood with respect to a trajectory~$\boldsymbol{T} = \left\{\xbf_j\right\}_{j=0,\ldots,M}$ with equidistant measurement intervals~$\tau$. 

Here we present the derivation of an approximate likelihood for such processes that avoids the biases discussed in the previous subsection. The derivation and final expression have parallels with Ref.~\cite{Ferretti2020} where models with a drift comprised of a linear velocity-dependent term and a non-linear position-dependent term were considered. In contrast to this previous work, here, we focus on models independent of position but with drift terms non-linear in velocity; generalizations to additional positional dependence are straightforward. Furthermore, the derivation and approximation presented here focuses on the often used secant velocities. Finally, we show how the final expression that contains a tridiagonal matrix like in Ref.~\cite{Ferretti2020} can be evaluated in linear time.  Details of the derivation  and intermediate calculations can be found in  Supplementary Note 2. 

First, we generalize the notion of \textit{secant velocity} to arbitrary points in time:
\begin{gather}
	\Vbf^{(\tau)}(t) \defeq \frac{1}{\tau} \int\limits_t^{t+\tau}\ddint{t'}\vbf(t') = \frac{\xbf(t+\tau) - \xbf(t)}{\tau}\,. \label{eq:trgau_integraltrafo}
\end{gather}
Formally, this time-continuous function~$\Vbf^{(\tau)}(t)$ is an integral transform of the instantaneous velocity~$\vbf(t)$ and can also be viewed as a smoothing with a box kernel of width~$\tau$ and height~$1/\tau$. We use the superscript~$(\tau)$ to stress that the definition of the time-continuous secant velocity depends on the measurement interval~$\tau$. 
Note that for~$t=t_j$ with~$j=0,1,\ldots M-1$, the value of the secant velocity can be obtained from the measured data. 

We apply the integral transform in Eq.~\eqref{eq:trgau_integraltrafo} to both sides of Eq.~\eqref{eq:trgau_SDEv} and obtain an SDE for the secant velocity 
\begin{gather}
	\dot{\Vbf}^{(\tau)}(t) = \fbf(\Vbf^{(\tau)}(t)) +\sqrt{2 D} \, \zetabf(t) + \mathcal{O}(\tau)\,.
	\label{eq:trgau_SDEV}
\end{gather}
Note that this SDE has the same form as Eq.~\eqref{eq:trgau_SDEv}, but is driven by colored noise~$\zetabf (t) \defeq \frac{1}{\tau} \int_t^{t+\tau}\ddint{t'}\xibf (t')$ with a finite correlation time instead of white noise. 
From now on, we will omit the superscript~${(\tau)}$ for better readability. 

Discretization of  Eq.~\eqref{eq:trgau_SDEV} yields
\begin{gather}
	\vec{Q}_j = \sqrt{2D}\int_{t_j}^{t_{j+1}}\ddint{t}\zetabf(t) + \mathcal{O}(\tau^{3/2})\,,
\end{gather}
with 
\begin{gather}
	\boldsymbol{Q}_j \defeq \Vbf_{j+1} - \Vbf_j - \frac{\tau}{2} \Big [ \fbf(\Vbf_{j+1}) +\fbf(\Vbf_{j}) \Big ] \,.
	\label{eq:trgau_transform}
\end{gather}
We approximate the joint probability density of $\vec{Q}_{0:M-2}$ by the Gaussian density of $\sqrt{2D}\int_{t_j}^{t_{j+1}}\zetabf(t)$:
\begin{gather}
	p(\vec{Q}_{0:M-2}) \approx \prod_\alpha \frac{\exp \left ( {-\frac{Q_i^\alpha \left[\mathbf{Z}^{-1}\right]_{ij} Q_j^\alpha}{4 D}  } \right )}{\sqrt{\det(4\pi D \mathbf{Z})}}
     \,,
	\label{eq:trgau_pQ}
\end{gather}
where we use Einstein's summation convention for indices~$i$ and~$j$.
$\mathbf{Z}$ is a matrix of the form:
\begin{gather}
	\mathbf{Z} = \frac{\tau}{6}
	\begin{pmatrix}
		4 & 1 &   &   &  \\
		1 & 4 & 1 &   &  \\
		& 1 & 4 & 1 &  \\
		&   & \ddots &\ddots&\ddots
	\end{pmatrix}.
	\label{eq:trGau_Z}
\end{gather}
The tridiagonal structure of~$\mathbf{Z}$ encodes the non-Markovianity of the position process, i.e., the additional correlations in~$\vec{V}_j$ that are absent for~$\vec{v}_j$ (compare also Ref.~\cite{Ferretti2020}). 

The vector $\vec{Q}_{0:M-2}$ is a (in general) non-linear function of the measured secant velocities $\vec{V}_{0:M-1}$.
In order to convert the probability of~$\boldsymbol{Q}_{0:M-2}$ in Eq.~\eqref{eq:trgau_pQ} into a probability~$p(\Vbf_{1:M-1}|\Vbf_0)$, we thus need to multiply the appropriate Jacobi determinant~$\mathcal{J}$ of the transformation. Because~$\vec{Q}_j$ only depends on~$\vec{V}_j$ and~$\vec{V}_{j+1}$, the determinant factorizes and we obtain
\begin{gather}
	\mathcal{J} = \prod_{j=1}^{M-1}\abs{\det \! \left ( \mathbb{1} - \frac{\tau}{2} \frac{\partial \fbf(\Vbf_j)}{\partial \Vbf_j} \right ) }.
%
    %
\end{gather}
 
The above considerations lead to the following approximation for the probability of the measured secant velocities:
\begin{widetext}
\begin{align}
	p(\Vbf_{0:M-1}|\etabf) &\approx \mathcal{J} p_{\vbf}(\Vbf_0|\etabf) \prod_\alpha\left[ \frac{1}{\sqrt{\det(4\pi D \mathbf{Z})}} \exp\mleft(-\frac{Q_i^\alpha \left[\mathbf{Z}^{-1}\right]_{ij} Q_j^\alpha}{4 D} \mright) \right] \,,
	\label{eq:trgau_fulllikelihood}
\end{align}
\end{widetext}
where again Einstein's summation convention is used for indices $i$ and $j$. 
Remember that the drift function~$\fbf$, and through it~$Q_i^\alpha$, depends on the motility parameters~$\boldsymbol{\eta}$.

Above, we have multiplied a probability for $\Vbf_0$ to obtain the joint probability of all measured secant velocities $\Vbf_{0:M-1}$. Since we assume no prior knowledge about $\Vbf_0$ we should include the steady state distribution. However, since, in general, it doesn't have an analytic expression, we use the steady state distribution of the instantaneous velocity~$p_{\vbf}$ as an approximation instead. For isotropic processes with $\fbf(\vbf) = f_v(v)\,\vbf$, the stationary Fokker-Planck equation is effectively one-dimensional and $p_{\vbf}$ is the (analytic or numerically approximated) solution to this one-dimensional PDE. For general processes the high-dimensional Fokker-Planck equation needs to be solved approximately or the steady-state distribution evaluated numerically. It is also possible to leave out the probability of $\Vbf_0$ which means ignoring any information contained in this term.

Taken as a function of the motility parameters~$\etabf$, the RHS of Eq.~\eqref{eq:trgau_fulllikelihood} is an approximation to the likelihood of the motility parameters with respect to the measured secant velocities. 
Because we assume the probability of the first measured position~$\xbf_0$ to be uniform over the observation space, the likelihood in terms of the positions is identical to the likelihood in terms of the secant velocities up to a~$\tau$ dependent constant:
\begin{align}
	L(\etabf) \defeq \log p(\boldsymbol{T}|\etabf) = \log p(\Vbf_{0:M-1}|\etabf) + C \,.
\end{align}
The expression for the likelihood in Eq.~\eqref{eq:trgau_fulllikelihood} is a transformation of the multivariate Gaussian distribution of~$\boldsymbol{Q}_{0:M-2}$ in Eq.~\eqref{eq:trgau_pQ}; therefore we will call this approximation to the likelihood \textit{transformed Gaussian} method. Note, however, that Eq.~\eqref{eq:trgau_fulllikelihood} is generally non-Gaussian with respect to the secant velocities due to the nonlinear drift~$\vec{f}(\Vbf)$.

Evaluation of the approximation to $p(\Vbf_{0:M-1}|\etabf)$ requires the calculation of the Gaussian density of~$\boldsymbol{Q}_{0:M-2}$ in square brackets in Eq.~\eqref{eq:trgau_fulllikelihood}.
The expression contains the inverse of $\mathbf{Z}$ as well as a double sum with $(M-1)^2$ terms. 
However, in the Methods section we show that the likelihood can still be efficiently evaluated in~$\mathcal{O}(M)$ steps due to the tridiagonal structure of $\mathbf{Z}$. This decreases the computational cost significantly.

\begin{figure}[htb]
	\includegraphics[width=\columnwidth]{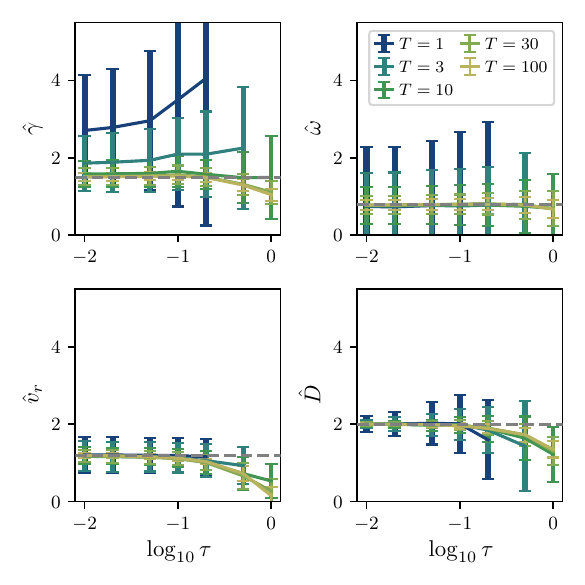}
    \vspace*{-0.85cm}
	\caption{\textbf{Maximum likelihood estimates for motility parameters of an examplary model}. The drift of the examplary model is given in~Eq.~\eqref{eq:trGau_ChMHatmodel_f} with drift parameters $\gamma, \omega, v_r$; $D$ is the parameter for the noise strength (see Eq.~\eqref{eq:trgau_SDEv}). The estimates were obtained by numerical maximization of the transformed Gaussian approximation of the likelihood. The plots show the mean of inferred parameters from 100 simulated trajectories; the errorbars show the corresponding standard deviations. The input parameter values of the simulation are shown as gray dashed lines. For different values of~$T$, the same set of trajectories was used and appropriately shortened. Trajectory simulation was done using an Euler-Maruyama scheme with a time step of $\min(\tau/100, 10^{-4})$.}
	\label{fig:trgau_consistency}
\end{figure}

While we focus on the case of observing the integral of a stochastic process, the above derivation also works for observations of general local means of a process, cf.~\cite{Gloter2008}: in this case, the box kernel in Eq.~\eqref{eq:trgau_integraltrafo} is replaced by some general local kernel. Note that in this case $\mathbf{Z}$ will have different entries.

\subheading{Consistent maximum likelihood estimates for motility parameters}
Before we move on to using the likelihood approximation Eq.~\eqref{eq:trgau_fulllikelihood} for heterogeneity estimation, we first illustrate its validity and effectiveness as an approximation to the single trajectory log-likelihood~$L(\boldsymbol{\eta})$. In order to do so, we use it to calculate MLEs of the motility parameters. 
The results are plotted in Fig.~\ref{fig:trgau_consistency} for a paradigmatic motility model for active particles in two spatial dimensions~\cite{Romanczuk2012,liebchen2022chiral} with the drift 
\begin{align}
	\vec{f}(\vec{v}) = -\gamma \boldsymbol{v} \! \left (v^2 - v_r^2 \right ) + \omega\!\begin{pmatrix}0&-1\\1&0\end{pmatrix} \vbf\,.
	\label{eq:trGau_ChMHatmodel_f}
\end{align}
Here,~$v=\abs{\vbf}$ denotes the speed of the particle. Together with the diffusion constant $D$, the motility parameters are~$\etabf=\left(\gamma, \omega, v_r, D\right)$. The drift term corresponds to a Mexican-hat potential with preferred speed~$v_r$. Additionally, there is a chiral term which makes the particles rotate. The plots show that the transformed Gaussian method for the approximation of the likelihood leads to consistent estimates of the motility parameters for sufficiently small measurement intervals~$\tau$. Only for estimation of the~$\gamma$ parameter there is effectively a minimum total recording time~$T = M\tau$ required to obtain consistent estimates---short trajectories contain little information about this parameter. 
As expected, the spread of the estimates for each of the parameters decreases with increasing trajectory length as the additional information enables more accurate inference.

%% file: EM.tex
\label{ssec:EM_algo}
The expression for the log-likelihood of the heterogeneity parameters in Eq.~\eqref{eq:MLE_likelihood} contains integrals over the motility parameters. 
In general, there is no analytic solution for these integrals. Instead of evaluating them numerically~\cite{Picchini2010}, we use an expectation-maximization~(EM) algorithm~\cite{Dempster1977} to maximize the likelihood in a similar way as in Ref.~\cite{DelattreLavielle2013}. 
The algorithm avoids evaluation of the integrals over the unobserved motility parameters~$\left\{\etabf^n\right\}$ and instead relies on calculation of expectation values. 

The EM algorithm is initialized by some initial guess for the heterogeneity parameters~$\hat{\thetabf}_0$. It then iteratively creates a sequence of estimates~$\{\hat{\thetabf}_i\}$ which converge to a local maximum of the log-likelihood $\mathcal{L}(\thetabf)$. This is achieved by the following iteration rule:
\begin{subequations}
	\label{eq:EM_EMfull}
\begin{align}
	\! \hat{\thetabf}_{i+1} &\!=\!\underset{\thetabf}{\mathrm{argmax}}\,\sum_n\int \! \dd{\etabf} p(\etabf|\boldsymbol{T}^n, \hat{\thetabf}_i)\log p(\boldsymbol{T}^n, \etabf|\thetabf) \! \\
	&\!=\! \underset{\thetabf}{\mathrm{argmax}}\,\sum_n \mathbb{E}_{n, i}\mleft[\log p_{\etabf}(\etabf|\thetabf)\mright]\,,
	\label{eq:EM_EM}
\end{align}
\end{subequations}
where~$\mathbb{E}_{n, i}[\cdot]$ denotes an expectation value with respect to the conditional probability distribution for the motility parameters
\begin{gather}
	p(\etabf|\boldsymbol{T}^n, \hat{\thetabf}_i)\sim p(\boldsymbol{T}^n|\etabf)\, p_{\etabf}(\etabf|\hat{\thetabf}_i)\,.
	\label{eq:EM_density_meth}
\end{gather}
The expectation values are usually not analytically tractable. An MCMC sampling scheme can be employed to draw samples proportional to $	p(\etabf|\boldsymbol{T}^n, \hat{\thetabf}_i)$ and approximate these expectations with sample averages. 

The procedure described above is, however, computationally expensive since we have to resample at each iteration step for each of the measured trajectories. In order to accelerate the calculations, we employ importance sampling~\cite[chap.\,3.3]{Robert2004}.  
We can create samples proportionally to each single trajectory likelihood~$\exp L_n(\etabf) \defeq p(\boldsymbol{T}^n|\etabf)$ once using the transformed Gaussian approximation~[Eq.~\eqref{eq:trgau_fulllikelihood}]. These samples can then be reused in each step of the EM algorithm with appropriate $\hat{\thetabf}_i$~dependent weighting factors to calculate expectation values with respect to $p(\etabf|\boldsymbol{T}^n, \hat{\thetabf}_i)$.

Care needs to be taken, because~$p(\boldsymbol{T}^n|\etabf)$ is not necessarily normalizable when taken as a function of~$\etabf$. By restricting the sampling space to some finite volume of the parameter space or multiplying a decaying function that ensures normalizability, we can still use the likelihood as a basis for sampling.

Details on the implementation of the EM algorithm can be found in Supplementary Note 5.

In the next section we combine the transformed Gaussian method for likelihood approximation with the EM algorithm to estimate the heterogeneity in synthetic datasets.

%% file: RES_CONST.tex
\begin{figure*}[htb]
	\includegraphics[width=\textwidth]{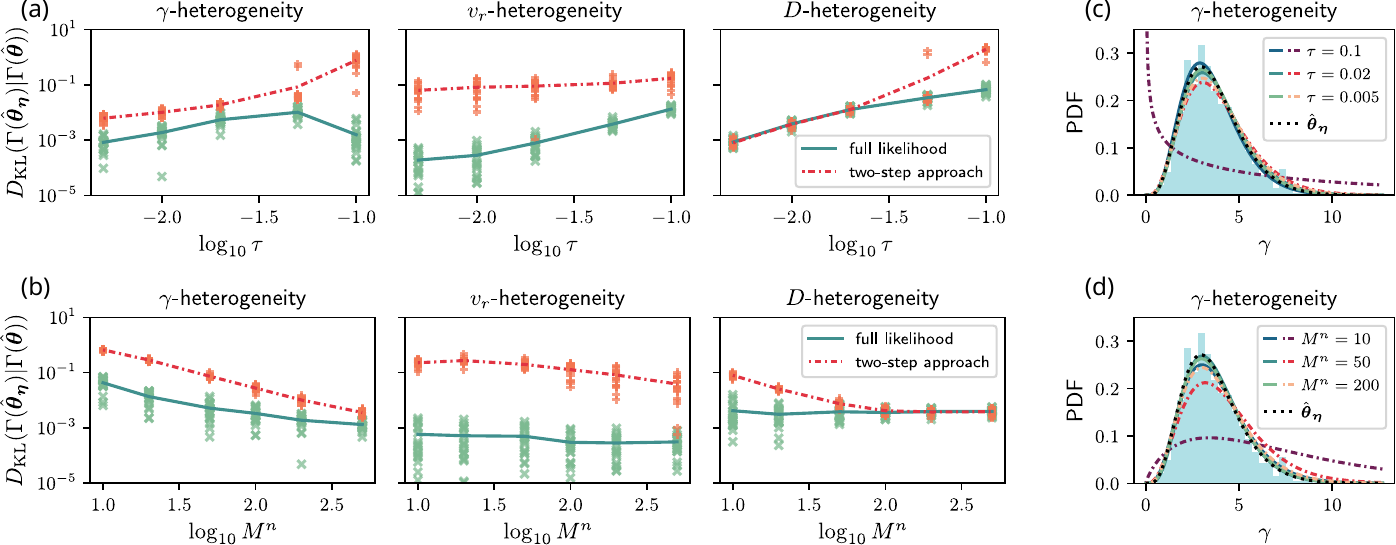}
	\vspace*{-0.5cm}
	\caption{\textbf{Comparison of heterogeneity inference with full likelihood and two-step approach.} The underlying exemplary motility and heterogeneity model are given in Eqs.~(\ref{eq:Consistency_motmodel},~\ref{eq:Consistency_hetmodel}); $\gamma$ and $v_r$ are the drift parameters, while $D$ controls the noise strength. Panels (a) and (b) show the KL-divergence between the heterogeneity distributions inferred using hidden motility parameters and inferred using the observed trajectories. The green cross symbols correspond to estimates obtained using the full likelihood~$\mathcal{L}(\thetabf)$, cf.~Eq.~\eqref{eq:MLE_likelihood}; the red plus symbols correspond to results for a two-step estimate where first the motility parameter is estimated for each trajectory and these estimates~$\left\{\hat{\etabf}\right\}$ are subsequently used to estimate the heterogeneity. In~(a) the dependence on sampling interval~$\tau$ for fixed~$T^n=2.0$ is shown; in~(b) the dependence on number of sample points~$M^n$ with fixed~$\tau=0.01$. For each parameter combination, the results of $R=20$ realizations of datasets with~$N=500$ trajectories are shown. The lines connect the arithmetic averages of the $R$ realizations. The input heterogeneity parameters are~$\thetabf = (\alpha_\gamma, \beta_\gamma, \alpha_{v_r}, \beta_{v_r}, \alpha_D, \beta_D)=(5.5, 1.5, 5.5, 1.5, 5.5, 0.5)$. Panels~(c) and~(d) show the inferred $\Gamma$-PDFs for individual realizations of a dataset as a function of~$\tau$ and~$M^n$, respectively. Solid lines correspond to full likelihood inference while dash-dotted lines correspond to the two-step approach. For reference, the histogram shows the true motility parameters used to generate the synthetic dataset. These are kept the same for all choices of~$\tau$ and~$M^n$.}
	\label{fig:Consistency_KLDiv}
\end{figure*}
The transformed Gaussian approximation in Eq.~\eqref{eq:trgau_fulllikelihood} can be used to approximate $p(\boldsymbol{T}^n|\etabf^n)$ in Eq.~\eqref{eq:MLE_likelihood}.
In this section, we demonstrate through numerical experiments that the likelihood based heterogeneity inference can indeed recover an input heterogeneity. 

In the setting that we are considering in this paper, the primary interest is finding the heterogeneity distribution~$p_{\boldsymbol{\eta}}(\boldsymbol{\eta}|\boldsymbol{\theta})$. 
In this sense the estimated heterogeneity parameters~$\hat{\thetabf}$ are merely an aid for finding the shape of the distribution. To judge the performance of the inference scheme, we therefore choose to use the Kullback-Leibler~(KL) divergence as a measure of similarity between distributions rather than the distance in $\thetabf$ space (see Methods section). 

We will compare the heterogeneity distribution inferred from the measured data, \ie, the trajectories, to the hypothetical case in which the hidden motility parameters~$\{\etabf_n\}$ are known. Given~$\{\etabf_n\}$, we use the maximum likelihood estimator~$\hat{\thetabf}_{\etabf} \defeq \mathrm{argmax}_{\thetabf} \sum_n \log p_{\etabf}(\etabf^n|\thetabf)$ 
to estimate the heterogeneity parameters. We use this as the baseline rather than the true input distribution because the trajectories cannot contain more information about the true heterogeneity distribution than the information contained in the hidden, finite set of motility parameters. In this sense,~$\hat{\thetabf}_{\etabf}$ represents the best possible guess from the given set of trajectories. 

In addition to inference by maximization of the full likelihood~$\mathcal{L}(\thetabf)$, cf.~Eq.~\eqref{eq:MLE_likelihood}, we also consider an alternative two-step approach. This approach first uses the transformed Gaussian approximation to obtain a single point MLE for the motility parameters~$\hat{\etabf}^n$ of each trajectory. 
In a second step we then use only these estimates to obtain an MLE of the heterogeneity parameter~$\hat{\thetabf} \defeq \mathrm{argmax}_{\thetabf} \sum_n \log p_{\etabf}(\hat{\etabf}^n|\thetabf)$.

As an example, we consider a two-dimensional system with drift function~\cite{Romanczuk2012} 
\begin{align}
	\vec{f}(\boldsymbol{v}) = -\gamma\, \boldsymbol{v} \left (v^2 - v_r^2 \right)
	\label{eq:Consistency_motmodel}
\end{align}
and heterogeneity model given by 
\begin{align}
	p_{\etabf}(\etabf|\thetabf) = q_{\Gamma}(\gamma|\alpha_\gamma, \beta_\gamma)\, q_{\Gamma}(v_r|\alpha_{v_r}, \beta_{v_r})\,
	q_{\Gamma}(D|\alpha_D, \beta_D),\label{eq:Consistency_hetmodel}
\end{align}
in which 
\begin{gather}
	    \label{eq:gampdf}
	q_{\Gamma}(x|\alpha, \beta) = \frac{\beta^\alpha}{\Gamma(\alpha)}x^{\alpha -1}e^{-\beta \,x}
\end{gather}
is the probability density function (PDF) of a~$\Gamma$-distribution $\Gamma(\alpha, \beta)$. Numerical results for further motility and heterogeneity models, including one with correlated motility parameters, can be found in Supplementary Note 6. The above drift is a version of Eq.~\eqref{eq:trGau_ChMHatmodel_f} with zero chirality~($\omega=0$). The model has three motility parameters~$\etabf=(\gamma, v_r, D)$ and a six dimensional heterogeneity parameter
\begin{gather}
	\thetabf = (\alpha_\gamma, \beta_\gamma, \alpha_{v_r}, \beta_{v_r}, \alpha_D, \beta_D)\,.
\end{gather}
The results for the inference of this model are shown in Fig.~\ref{fig:Consistency_KLDiv}. We can see that our inference scheme does indeed converge to the baseline distribution for sufficiently long trajectories and small sampling intervals $\tau$.

How accurately the heterogeneity of each of the parameters can be inferred depends on three factors. First, any approximation errors in the single trajectory likelihood will lead to biases in the heterogeneity inference. As discussed above, the single trajectory likelihood approximation in Eq.~\eqref{eq:trgau_fulllikelihood} becomes accurate in the limit of small measurement intervals $\tau$. In Fig.~\ref{fig:Consistency_KLDiv}(a) we see that the inferred distributions for all parameters do indeed approach the baseline for decreasing $\tau$. This is true for our likelihood approach as well as the alternative two-step approach since the latter one also uses the approximate likelihood to obtain the MLEs for the motility parameters. 	

Second, the width of the single trajectory likelihood indicates how much (Fisher) information about the motility parameters is contained in each trajectory. The more information per trajectory, the better the distribution of the motility parameters can be inferred. The information about the drift parameters depends on the duration of the trajectory~\cite{FrishmanRonceray2020}. As expected the accuracy of the inference for the $\gamma$-heterogeneity in Fig.~\ref{fig:Consistency_KLDiv}(b) increases steadily as $M^n$, and with it $T^n$, increases. The effect is weaker for the $v_r$ heterogeneity. This is most likely because the stationary distribution is more sensitive on changes in $v_r$ than  changes in $\gamma$. Information about the diffusion parameter, on the other hand, depends mostly on the sampling interval. Once a certain small number of data points have been observed, additional measurements only marginally decrease the width of the likelihood in $D$ direction. Therefore, the improvement in inference with larger $M^n$ in Fig.~\ref{fig:Consistency_KLDiv}(b) quickly stagnates for the $D$-heterogeneity.

The width of the single trajectory likelihood (relative to the width of the heterogeneity) is also responsible for the differences between the full likelihood approach and the alternative two-step approach. 
Inference from the full likelihood implicitly takes into account all possible values for the motility parameters and weights them according to the single trajectory likelihood; the two-step approach on the other hand only considers a single value of the motility parameter for each trajectory. If there is a lot of information, the single trajectory likelihood is sharp and the weighting function approaches a delta distribution. In this case the inferred heterogeneity from the full likelihood will be almost identical to the inference form the single point estimates. This is the case for $D$ for many data points and/or short sampling intervals, as well as for $\gamma$ in the limit of long trajectory durations. On the other hand, little information per trajectory leads to large deviations between the two approaches as can be seen for the drift parameters at short trajectory durations [cf.~Fig.~\ref{fig:Consistency_KLDiv}(b),(d)]. 

Finally, the quality of inference of course also depends on the number of observed trajectories. The results in Fig.~\ref{fig:Consistency_KLDiv} are obtained for a large number of trajectories~($N=500$). Since we choose $\hat{\thetabf}_{\etabf}$ as a baseline, those results will only depend weakly on $N$, though. 

\begin{figure}[htb]
	\includegraphics[width=\columnwidth]{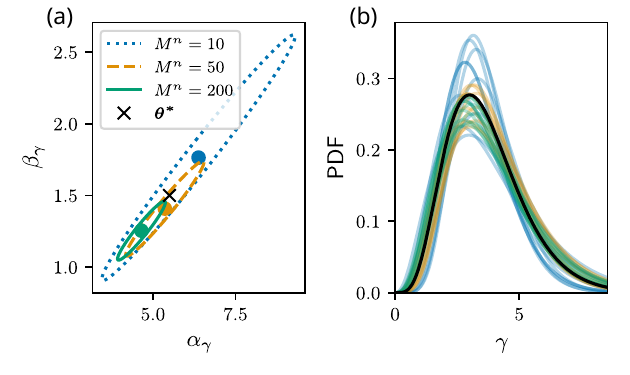}
	\caption{\textbf{Exemplary uncertainty bounds of heterogeneity estimates.} Results for the $\gamma$-parameter heterogeneity of the model given in Eqs.~(\ref{eq:Consistency_motmodel},~\ref{eq:Consistency_hetmodel}) from datasets with $N=100$ trajectories are shown. The different colors correspond to three datasets with different trajectory durations. The same heterogeneity parameters as in Fig.~\ref{fig:Consistency_KLDiv} were used; the sampling interval was $\tau=0.01$. Plot (a) shows the heterogeneity estimates together with the 1-$\sigma$ ellipses as given by the Hessian matrix in heterogeneity space. $\alpha_{\gamma}$ and $\beta_{\gamma}$ are the parameters of the $\Gamma$-distribution which is used as the heterogeneity model (see Eq.~\eqref{eq:Consistency_hetmodel}). In plot (b), density functions corresponding to several points along these ellipses are shown that visually create an uncertainty ``sleave''. The black cross in (a) and the black line in (b) correspond to the input heterogeneity.}
	\label{fig:Consistency_uncertainty}
\end{figure}

The curvature of the likelihood $\mathcal{L}(\thetabf)$ around its maximum provides information about the expected deviation of the MLE from the true value. If the curvature is large, other possible values of $\thetabf$ are unlikely; if it is small, \ie, the likelihood is relatively flat around the maximum, there is a wide range of other likely values. The Hessian matrix (see Eq.~\eqref{eq:MLE_hessianmatrix}) is a measure of this curvature and is directly connected to the expected variance of the estimator for repeated realizations of the same experiment (see Methods section and Supplementary Note 4). It therefore provides uncertainty estimates for the MLE $\hat{\thetabf}$. Conveniently, the Hessian matrix can be approximately calculated using the same set of samples that have already been used in the EM algorithm. Figure \ref{fig:Consistency_uncertainty} shows exemplary uncertainty bounds derived from the Hessian matrix for three datasets with different trajectory durations. As expected, the uncertainty increases for shorter durations. Note that the uncertainty bounds in Fig.~\ref{fig:Consistency_uncertainty} are tilted since $\alpha$ and $\beta$ are not orthogonal parameters. Orthogonal parameters of the $\Gamma$ distribution can be derived by diagonalization of the corresponding Fisher information matrix \cite{BarndorffNielsen1994}. Further numerical evidence for the robustness of these uncertainty bounds can be found in Supplementary Note 4.

%% file: CONC.tex
We proposed a methodology to estimate the heterogeneity of a population of active particles from discrete trajectory data. We opted for a full likelihood approach as this utilizes the available data efficiently, allows for consistent uncertainty measures of the estimate, and outperforms alternative two-step approaches especially in the limit of short trajectories. 
We show that the approach successfully separates the two sources of stochasticity in the data: the temporal fluctuations of the trajectories and the variability between the individuals. This way, it opens the door to quantitative modeling of heterogeneous motile particle ensembles. 

At the heart of the method is a closed-form approximation of the single trajectory likelihood for second-order Langevin models. The second-order nature of the model makes the observed processes non-Markovian, which requires more involved approximation techniques than in the case of first-order models. Our approximation approach is in the spirit of the \textit{first integrate then discretize} approach in Ref.~\cite{Ferretti2020}. 
We considered finite difference approximations to the instantaneous velocities, which we call secant velocities. These are effectively driven by colored noise, so they have different correlations from the instantaneous velocities, which are driven by white noise. A Gaussian distribution with a tridiagonal correlation matrix accounts for that colored noise and enables a consistent approximation of the likelihood in the limit of small sampling intervals. In contrast to filtering approaches~\cite{DelattreLavielle2013}, this method yields a compact functional form of the likelihood while having the same~$\mathcal{O}(M)$ computational complexity. The likelihood approximation also works for general local means instead of secant velocities.

Samples drawn from a probability proportional to the single trajectory likelihoods can be used in an EM algorithm to obtain a maximum likelihood estimate for the parameters of the heterogeneity distribution. Using the same set of samples, the Hessian matrix of the heterogeneity likelihood can be approximated which provides an uncertainty measure that captures the variability of the estimator under different realizations.

We note that the inferred heterogeneity can also be used to improve the estimates of the individual motility parameters by using it as a prior in a maximum a  posteriori scheme: $\log p(\etabf|\boldsymbol{T}^n) = L_n(\etabf) + \log p_{\etabf}(\etabf|\hat{\thetabf}) + \mathrm{const.}$

In practice the likelihood-based heterogeneity inference as described in this paper will be part of a thorough statistical analysis of an experimental dataset. When presented with a new dataset of trajectories, researchers will first analyze statistical quantities like mean squared-displacement, velocity correlations and conditioned expectations to find an appropriate candidate for a motility model \cite{Grossmann2024, PedersenFlyvbjerg2016, Bodeker2010, Cherstvy2018}. With a motility model at hand, trajectory-wise parameter estimates in the spirit of the two-step approach give an idea if the population is heterogeneous and which type of heterogeneity model might fit the distribution. Once this is done, the likelihood-based inference provides accurate estimates with corresponding uncertainties. In our related publication Ref.~\cite{Albrecht2026Transport}, we apply this pipeline to an experimental dataset, where the resulting model leads to an analytically tractable likelihood of the heterogeneity parameters. Furthermore, Bayesian model comparison methods can be used on top, to compare and rate different candidate models.

The applicability of the inference method presented in this paper relies on a number of assumptions. We discuss them in the following and point at possible generalizations that loosen these assumptions.

First, we have assumed a constant time~$\tau$ between measurements. In experimental settings individual measurements are sometimes not recorded correctly resulting in  missing data points in trajectories. This leads to a time between measurements that is a multiple of~$\tau$. Our method can be adapted to these cases:~the missing data can be included in the sampling process of the motility parameters and subsequently marginalized, such that the motility parameters are effectively sampled from the likelihood with respect to the available data points.
For cases with irregular measurements, there is no timescale~$\tau$ and therefore a secant velocity cannot be defined consistently across a trajectory. However, it should still be possible to set up an equivalent of Eq.~\eqref{eq:trgau_pQ} where the covariance matrix has a more complicated structure which depends on the durations between the measurements.
The potential advantages of such an approach compared to extended Kalman filters~\cite{DelattreLavielle2013, Sarkka2013} remain to be evaluated. 

Furthermore, we assumed that there is no measurement error and positions can be measured exactly. This is a valid approximation e.g.~for observations of cells with amoeboid motility~\cite{Cherstvy2018} where the area center of the cell can be calculated with sub-pixel accuracy. In other scenarios, it might be necessary to take the measurement uncertainties into account. In these cases, the measurements become 
\begin{gather}
	\vec{y}_j = \xbf_j + \vec{\varepsilon}_j\,,
\end{gather}
with~$\vec{\varepsilon}_j$ representing the measurement noise. The measured noisy secant velocities will then take the form 
\begin{gather}
	\bm{\mathcal{V}}_j = \frac{\bm{y}_{j+1} - \bm{y}_j}{\tau} = \Vbf_j + \frac{\vec{\nu}_j}{\tau}\,,\label{eq:conc_measuredsecant}
\end{gather}
where~$\vec{\nu}_j \defeq \vec{\varepsilon}_{j+1} - \bm{\varepsilon}_j$. 
Note that the noise terms~$\vec{\nu}_j$ in Eq.~\eqref{eq:conc_measuredsecant} are correlated with their neighbors. 
The single trajectory likelihood with respect to the measured secant velocities is given as
\begin{gather}
	p(\bm{\mathcal{V}}_{0:M-1}|\etabf) = \!\! \int \! (\mbox{d} {\Vbf})^{\!M} p(\bm{\mathcal{V}}_{0:M-1}|\Vbf_{0:M-1})\,p(\Vbf_{0:M-1}|\etabf)\,.
\end{gather}
The integrals above are generally untractable analytically, but could be approached using standard schemes, e.g.~Laplace approximations. It remains to be studied if measurement uncertainty can be approximately treated in the sequential evaluation of the Gaussian density as presented in the Methods section.

In this paper, we have considered drift functions that depend on velocity only. Additional position dependence of the drift can be included in the derivation of the transformed Gaussian method without altering the scaling behavior of the correction term. Note that in this case an additional term for the steady state distribution of the initial position~$\xbf_0$ is needed. 

Extensions to models with noise multiplicative in the velocity are not readily possible under the current framework, though. In the drift in Eq.~\eqref{eq:trgau_SDEV}, the instantaneous velocity could just be replaced by the secant velocity which only introduced a~$\mathcal{O}(\tau)$ correction. Doing the same with a multiplicative noise term would lead to correction terms larger than~$\mathcal{O}(\tau)$. It remains to be studied if the additional correction terms or more intricate modifications still allow for consistent likelihood approximations in the spirit of the transformed Gaussian method. Alternative methods to tackle models with multiplicative noise are Gaussian filter approaches~\cite{DelattreLavielle2013} possibly enhanced by Taylor moment expansions~\cite{Zhao2021} or using full blown particle methods~\cite{Whitaker2017, Wiqvist2021}. 

The EM algorithm that we use to maximize the likelihood of the heterogeneity parameters relies on sample averages with samples drawn from a probability proportional to the single trajectory likelihood. For large numbers of motility parameters or complicated multi-modal likelihoods the sampling scheme used here (see Methods section) might not be sufficient. More sophisticated sampling schemes exchange Monte Carlo methods~\cite{Hukushima1996} can be employed in these cases.

Finally, the approach assumes constant motility parameters along the trajectory. However, since the full likelihood approach is well suited to deal with information scarce situations, short snippets of trajectories can be analyzed in which the motility parameters can be assumed to be constant. This way, non-stationary effects in datasets can be uncovered \cite{Albrecht2026Transport}.

We see our approach as a step on the way to analyzing the variability and heterogeneity of motility patterns more thoroughly. 
While we have considered MLEs in this work, the method can readily be used in Bayesian frameworks. 
Provided with a prior on~$\thetabf$, the EM algorithm can be modified to obtain a maximum posterior estimate~\cite[chap. 9.4]{Bishop2006}. Additionally, the transformed Gaussian method can also be used in sampling schemes that draw samples from the posterior distribution. 
When applied to real data, the method can be augmented by model comparison techniques in order to find the models that best describe the given system. A further future direction is to include interaction terms in the motility model in order to be able to extract heterogeneity within collective motion patterns~\cite{ariel2015order,Castro2021,supekar_learning_2023}, the response to external fields~\cite{rode2024information} and particle-wall interactions~\cite{Lambert2025}. Finally, we note that our approach is not restricted to motility data. The method for likelihood calculation can in general be applied to systems in which a smoothed out version of the input data is observed, as for example in ice-core data~\cite{DitlevsenDitlevsen2002}.

%% file: HESS_MAT.tex
\subheading{Uncertainty estimate via Hessian Matrix}
The EM algorithm yields an approximate maximum likelihood estimator~$\hat{\thetabf}$ for the heterogeneity parameters. 
In order to judge the reliability of this estimate, we want to estimate the uncertainty of~$\hat{\thetabf}$. 
This can be done by calculating the Hessian matrix~$\mathbf{H}(\hat{\thetabf})$ [cf.~Eq.~\eqref{eq:MLE_hessianmatrix}], \ie, the second derivative, of the log-likelihood~$\mathcal{L}(\thetabf)$ at~$\hat{\thetabf}$.

First, if the likelihood is assumed to be a non-normalized Gaussian distribution, the negative inverse of the Hessian matrix corresponds to correlation matrix of this distribution. From a Bayesian perspective, the Hessian matrix therefore determines where the majority of the posterior mass is concentrated given a flat prior. Second, the negative Hessian matrix is an empirical estimate of the Fisher information~$\mathcal{I}(\vec{\theta}^*)$ about the heterogeneity parameters contained in the given set of trajectories. Typically, the fluctuations of an MLE around the true value converge in distribution to a normal distribution~\cite[chap. 7.3]{Schervish1995}:
\begin{gather}
	\sqrt{N}(\hat{\thetabf} -\thetabf^*)\overset{\mathcal{D}}{\rightarrow} \mathcal{N}(0, \mathcal{I}^{-1}(\thetabf^*))\qq{as} N\rightarrow\infty\, .
	\label{eq:UncEst_asympnorm}
\end{gather}
This property is called asymptotic normality and implies that the stochastic fluctuations of the MLE~$\hat{\vec{\theta}}$ are directly connected to the Fisher information in the limit of many trajectories in the dataset (see also Supplementary Note 4). 
This means that we can use the negative inverse Hessian matrix~$\mathbf{H}$ to obtain an estimate of the standard deviation of~$\hat{\thetabf}$.

After a few manipulations (see Supplementary Note 3), we find that the elements of the Hessian matrix can be expressed as
\begin{align}
	H_{\alpha\beta} = \sum_n &\left[\mathbb{E}_n\mleft[\frac{\partial_\alpha\partial_\beta p_{\etabf} (\etabf|\thetabf)}{p_{\etabf} (\etabf|\thetabf)}\mright]\right. \\
	&\;\left.\left.- \mathbb{E}_n\mleft[\frac{\partial_{\alpha}p_{\etabf} (\etabf|\thetabf)}{p_{\etabf}(\etabf|\thetabf)}\mright]\mathbb{E}_n\mleft[\frac{\partial_{\beta} p_{\etabf}(\etabf|\thetabf)}{p_{\etabf}(\etabf|\thetabf)}\mright]\right]\right|_{\thetabf = \hat{\thetabf}}\,, \nonumber
\end{align}
where~$\mathbb{E}_n[\cdot]$ denotes an expectation with respect to~$p(\etabf|\boldsymbol{T}^n, \hat{\thetabf})$ and we have used the abbreviation~$\partial_{\alpha}\defeq \partial/\partial \theta^{\alpha}$. This turns out to be a slightly modified version of Louis' formula~\cite{Louis1982, DelattreLavielle2013}. 
Conveniently, the expectations can be approximated by sample averages using the same set of samples that have been used in the EM algorithm.

%% file: APP.tex
\subheading{Sequential calculation of Gaussian density of~$\vec{Q}_{0:M-2}$}
\label{sec:APP_seqcalc}
The transformed Gaussian method to calculate the likelihood~[Eq.~\eqref{eq:trgau_fulllikelihood}] includes the evaluation of the multivariate Gaussian density approximation of~$Q^\alpha_{0:M-2}$~[Eq.~\eqref{eq:trgau_pQ}]. For simplicity, we consider here a rescaled one-dimensional version of the PDF,
\begin{gather}
	p(\chi_{0:M-2}) \approx  e^{-\frac{1}{2} \chi_i \left[\mathbf{Z}^{-1}\right]_{ij} \chi_j } \det(2\pi\, \mathbf{Z})^{-1/2}\,,
\end{gather}
where we used Einstein's summation convention with indices~$i, j=0, \ldots, M-2$. Due to indices starting with~$0$ in the main text, we stick to this convention here. Note that this means that in the following the highest index in a~$j\times j$ matrix will be~$j-1$ as in the programming languages C or Python.
Even though the matrix~$\mathbf{Z}$, which is given in Eq.~\eqref{eq:trGau_Z}, has an explicit inversion formula~\cite{Jia2013}, a straightforward evaluation of the above probability density still requires~$\mathcal{O}(M^2)$ calculations due to the double sum in the exponent. In this section we present a sequential method which only takes~$\mathcal{O}(M)$ calculations, thus dramatically decreasing the computational complexity.

We first note that the joint probability of~$\chi_{0:M-2}$ can be written as conditioned probabilities in the following way:
\begin{gather}
	p(\chi_{0:M-2}) = p(\chi_0) \, \prod_{j=0}^{M-3}p(\chi_{j+1}|\chi_{0:j})\,.
\end{gather}
Since the joint probability is Gaussian, the conditioned probabilities are as well:
\begin{gather}
	p(\chi_j|\chi_{0:j-1})= \mathcal{N}(\mu_j, \sigma^2_j)\,.
\end{gather} 
The means and variances of these densities can be calculated sequentially as we shall see. We let~$\mathbf{Z}_{j\times j}$ denote the~$j\times j$ version of~$\mathbf{Z}$. Using the rules for partitioned Gaussians~(see Ref.~\cite[chap.~2]{Bishop2006}), we find that
\begin{align}
	\mu_j =& \expval{\chi_j} + \left[\mathbf{Z}_{j+1\times j+1}\right]_{jk}\left[\mathbf{Z}^{-1}_{j\times j}\right]_{kl}(\chi_l - \expval{\chi_l})\nonumber\\
	=&\left[\mathbf{Z}_{j+1\times j+1}\right]_{j\,j-1}\left[\mathbf{Z}^{-1}_{j\times j}\right]_{j-1\,l}\chi_l\,,
\end{align}
where the Einstein sums run over indices~$k,l=0,\ldots, j-1$, and
\begin{gather}
	\sigma^2_j = \left(\left[\mathbf{Z}^{-1}_{j+1\times j+1}\right]_{jj}\right)^{-1}\,.
\end{gather}
According to Ref.~\cite{Jia2013}, the elements of the inverse of a~$j\times j$ matrix~$\mathbf{Z}_{j\times j}$ are
\begin{gather}
\left[\mathbf{Z}_{j\times j}^{-1}\right]_{kl} = \frac{6}{\tau} \left(\omega_\abs{k-l} - \omega_{\abs{k+l+2}}\right)
	\shortintertext{with}
	\omega_m = \frac{\beta^{m+1} + \beta^{2j-m + 3}}{(\beta^2 - 1)(1 - \beta^{2(j+1)})} , 
\end{gather}
where~$\beta = -2 + \sqrt{3}$ and ~$i,j=0,\ldots, k-1$. With this, we find that the variances are
\begin{gather}
	\sigma^2_j = \frac{\tau}{6}\left(-\beta\right)^{-1}\frac{1 - \beta^{2(j+2)}}{1 - \beta^{2(j+1)}}\,.
	\label{eq:App_sig2}
\end{gather}
The mean of the conditioned probabilities are given by
\begin{gather}
	\mu_j = \sum_{l=0}^{j-1}\frac{\beta^{j+l+2} - \beta^{j-l}}{1 - \beta^{2(j+1)}} \, \chi_l
\end{gather}
from which it follows that 
\begin{gather}
	\mu_{j+1} = \frac{1 - \beta^{2(j+1)}}{1 - \beta^{2(j+2)}}\, \beta \left(\mu_j - \chi_j\right).
\label{eq:App_musequential}
\end{gather}
Therefore, using Eqs.~(\ref{eq:App_sig2}, \ref{eq:App_musequential}) starting from~$\mu_0=0$, we can indeed calculate~$p(\chi_{0:M-2})$ in~$\mathcal{O}(M)$ steps. In order to calculate~$p(Q_{0:M-2})$, we only need to replace~$\tau / 6 \rightarrow 2 D \tau / 6$ in Eq.~\eqref{eq:App_sig2}.

\subheading{Kullback-Leibler divergence}
The Kullback-Leibler (KL) divergence between two distributions $P, Q$ with PDFs $p(x), q(x)$ is given by (see e.g.~\cite{Bishop2006})
\begin{gather}
	D_{\mathrm{KL}}(P|Q) = \int\ddint{x} p(x) \log \frac{p(x)}{q(x)}\,.
\end{gather}
It is non-negative with $D_{\mathrm{KL}}(P|Q) = 0$ if and only if $p(x)=q(x)$. By definition, the KL-divergence is the expected difference in logarithms between $p(x)$ and $q(x)$ and can therefore be used as a measure of dissimilarity between two distributions. Note that the KL-divergence is not symmetric: $D_{\mathrm{KL}}(Q|P)\neq D_{\mathrm{KL}}(P|Q)$. For the results shown in this paper, we use the reference distribution as the first argument and the inferred distribution as the second. This means that the expectation is taken with respect to the reference distribution.

The KL divergence between two $\Gamma$-distributions $\Gamma(\alpha_1, \beta_1)$ and $\Gamma(\alpha_2, \beta_2)$ with parametrization as in \eqref{eq:gampdf} is given by
\begin{align}
	D_{\mathrm{KL}}(\Gamma(\thetabf_1)|\Gamma(\thetabf_2)) =
	 &\log\mleft(\frac{\Gamma(\alpha_2)}{\Gamma(\alpha_1)}\mright) 
	+ \alpha_2 \log\mleft(\frac{\beta_1}{\beta_2}\mright)\nonumber \\
	&-\frac{\alpha_1}{\beta_1}(\beta_1 - \beta_2)
	+ (\alpha_1 - \alpha_2) \psi(\alpha_1)\,,
\end{align}
where $\Gamma(x)$ is the gamma function and $\psi(x) = \partial_x \log\Gamma(x)$ is the digamma function.

%% file: ACKN.tex
\noindent\textbf{Data availability}\\
The numerical data are available from the authors or can be generated independently using the provided computer code (see Code availability below). Correspondence and requests for materials should be addressed to JA or RG.

\noindent\textbf{Code availability}\\
The computer codes used for simulations and numerical calculations is available at \url{https://gitup.uni-potsdam.de/Biological_Physics_Group/likelihood-heterogeneity-inference}.

\noindent\textbf{Author contributions}\\
JA and RG developed the likelihood approximation. JA performed the numerical simulations and wrote the manuscript. RG and MO supervised the project. All authors discussed the results and contributed to the final manuscript.

\noindent\textbf{Funding Statement}\\
This research has been partially funded by the Deut\-sche For\-schungs\-ge\-mein\-schaft(DFG) -- Project-ID 318763901 -- SFB1294.

\noindent\textbf{Competing interests}\\
The authors declare no competing interests.